# Stretching-induced conductance variations as fingerprints of contact configurations in single-molecule junctions

Yong-Hoon Kim,[1,*] Hu Sung Kim,[1] Juho Lee,[1] Makusu Tsutsui,[2,*] and Tomoji Kawai[2]

[1] Graduate School of Energy, Environment, Water, and Sustainability, Korea Advanced Institute of Science and Technology, 291 Daehak-ro, Yuseong-gu, Daejeon 305-701, Korea.

[2] The Institute of Scientific and Industrial Research, Osaka University, 8-1 Mihogaoka, Ibaraki, Osaka 567-0047, Japan.

Molecule-electrode contact atomic structures are a critical factor that characterizes molecular devices, but their precise understanding and control still remain elusive. Based on combined first-principles calculations and single-molecule break junction experiments, we herein establish that the conductance of alkanedithiolate junctions can both increase and decrease with mechanical stretching and the specific trend is determined by the S-Au linkage coordination number (CN) or the molecule-electrode contact atomic structure. Specifically, we find that the mechanical pulling results in the conductance increase for the junctions based on S-Au CN two and CN three contacts, while the conductance is minimally affected by stretching for junctions with the CN one contact and decreases upon the formation of Au monoatomic chains. Detailed analysis unravels the mechanisms involving the competition between the stretching-induced upshift of the highest occupied molecular orbital-related states toward the Fermi level of electrodes and the deterioration of molecule-electrode electronic couplings in different contact CN cases. Moreover, we experimentally find a higher chance to observe the conductance enhancement mode under a faster elongation speed, which is explained by *ab initio* molecular dynamics simulations that reveal an important role of thermal fluctuations in aiding deformations of contacts into low-coordination configurations that include monoatomic Au chains. Pointing out the insufficiency in previous notions of associating peak values in conductance histograms with specific contact atomic structures, this work resolves the controversy on the origins of ubiquitous multiple conductance peaks in S-Au-based single-molecule junctions.

## Introduction

A major breakthrough in the molecular electronics research over the past decade was the introduction and refinement of scanning tunneling microscopy or mechanically controllable break junction methods to establish single-molecule junctions in a systematic and robust fashion.[1-2] Repeated formation of single-molecule junctions accompanied with the statistical data analysis has not only significantly improved the quantitative agreement among different experimental data but also continues to elucidate scientifically important yet complex phenomena that can potentially lead to novel applications.[2-3] The characteristic feature of these molecular junctions is that the molecule-electrode interfaces often play a role as much important as the molecules themselves.[4-5] As a representative manifestation, multiple energetically favorable contact atomic structures were claimed to produce separate conductance peaks.[6-19] While this feature could be ideally exploited as a potential route to realize a switching device,[13] it demonstrates the inherent variability and controllability issues in the single-molecule junction platform.[4-5] Particularly, it is still unfortunately the case that the correlation between contact atomic structures and charge transport characteristics is not fully established for the ubiquitous gold-surfur contacts,[20-21] which represent the most well-established venue toward molecular self-assembly and nanofabrication for the electronic, energy, and bio applications.[22-24]

Applying a combined computational and experimental approach, herein we show that the conductance variations during the mechanical stretching represents a unique fingerprint of contact atomic structures in covalent thiolate-gold (RS-Au) bond-mediated single molecule junctions. We particularly note a recent experiment[21] that has successfully observed the theoretically-predicted[25-26] conductance increase in pulled benzendithiolate (BDT) junctions. The counterintuitive observation was explained by the upshift of the BDT highest-occupied molecular orbital (HOMO) and related states toward the Fermi level ($E_F$) of Au electrodes upon mechanical stretching and the resulting enhancement of resonant transmission. However, questions still remain on why the conductance decrease rather than increase with junction elongation is more commonly observed irrespective of the HOMO upshift, and whether such behavior is general to all Au-S-based single-molecule junctions. In fact, the energetic positions of benzene molecular core and thiol contacts are comparable and their interplay can complicate the charge transport process throughout the molecular junctions.[17, 27-28] In this work, we instead employ the hexaneditholate (C6DT) molecule where the hexane molecular core energy levels are much farther away from $E_F$ and thus the role of Au-S contacts is maximized.[27, 29]

We experimentally show that the C6DT junction can exhibit the anomalous conductance increase upon junction pulling like the BDT counterpart, which establishes the generality of such behavior. Our theoretical analysis then identifies that the stretching-induced conductance variation trend is strongly correlated with the S-Au coordination number (CN), and the pulling-induced conductance increase appears only for the CN two (CN2) and CN three (CN3) cases. On the other hand, for the CN one (CN1) and related yet distinctive Au monoatomic chain-based contacts, which should be more probable configurations in break junction experiments,[30-33] we find that the junction stretching produces constant and decreasing conductance curves, respectively. This classification is justified by experiments performed at two different pulling rates and the corresponding *ab initio* molecular dynamics simulations of restructuring and bond breaking processes at the stretched Au-S contacts. The implications of our findings will be discussed in view of the multiple conductance peaks observed in single-molecule junction experiments and self-assembled monolayer junctions that are more relevant for practical device applications.

## Methods

**Single-molecule conductance measurements.** Displacement-dependent charge transport through Au-1,6-C6DT-Au junctions were explored at room temperatures in vacuum by using micro-fabricated mechanically-controllable break junctions. Specifically, we first broke a junction in a toluene solution of C6DT of concentration 1 μM by deflecting a phosphor bronze substrate with a piezo-actuator. This allowed C6DT molecules to adhere on the fleshly exposed fracture surface via strong Au-S bonding. Subsequently, we started to evacuate the chamber and removed the solvent so as to prevent molecular aggregation on the junction. The break junction experiments were then carried out in a vacuum better than $10^{-5}$ Torr by using a Keithley 6487 picoammeter-source unit to record the conductance $G$ under a constant dc bias voltage $V_b = 0.2$ V. Here, the control of contact mechanics was implemented with a conductance feedback loop wherein the junction stretching speed $v_d$ was adjusted to predefined values in the course of contact opening and closing processes.[34-35] Precisely, $v_d$ was set to 6 pm/s or 0.6 pm/s when $G$ was lower than 6 $G_0$.

**First-principles calculations.** In the static junction pulling simulations, we carried out density functional theory (DFT) calculations using the SeqQuest code (Sandia National Lab) within the local density approximation (LDA).[36] The atomic cores were replaced by norm-conserving pseudopotentials [37]. Gaussian basis sets of the double-zeta-plus-polarization quality optimized for the corresponding LDA pseudopotentials were adopted. In constructing junction models, five Au(100) and Au(111) layers were used to model each electrode and the vacuum gap of 20 Å was placed along the electrode surface-normal direction to make the interactions of slab models with their periodic images negligible. A single $\vec{k}_\parallel$-point shifted off from the $\Gamma$ point was sampled along the electrode-surface directions. Geometry optimizations were carried out until the total residual force is below $10^{-5}$ eV/ Å.

## Results and Discussion

**Experimental observation of conductance increase (CI) as well as the flat (FL) conductance variation and conductance decrease (CD) with mechanical stretching in C6DT junctions.** First, we discuss our experimental data by using micro-fabricated mechanically-controllable break junctions (Figure 1a) that show the conductance increase as well as conductance decrease upon junction elongation. Conductance versus time ($G$ - $t$) curves displayed stepwise decrease in $G$ during mechanical thinning of nanoscale Au junctions. Long plateaus were found at near 1 $G_0$ signifying formation of Au single-atom chains ($G_0$=2$e^2$/$h$ is the conductance quantum where $e$ and $h$ are the electron charge and Planck's constant, respectively).[45-47] In some cases, $G$ decreased sharply to zero after when the single-atom chains were fractured, which suggests quantum tunneling of electric charges across the vacuum gap between the closely separated Au nanoelectrodes formed after the breakdown of a contact. In other cases, the conductance traces revealed additional plateaus at below 1 $G_0$ associated with formation of Au-C6DT-Au structures. This characteristic feature can be found in the two-dimensional histogram (constructed with more than 1800 $G$-$t$ curves obtained at $v_d$ = 6 pm/s without any selection) showing data clustered at around 1 $G_0$ attributed to the electron transport characteristics of Au

single-atom contacts and also in a low-$G$ regime from $10^{-3}$ $G_0$ to $10^{-4}$ $G_0$, the range of which covers the conductance of Au-C6DT-Au single-molecule junctions reported in the previous studies (Figure 1b).[6, 48] The fact that most junctions are dissociated under relatively small amount of displacement, $L_d \sim 1$ nm, indicates the non-negligible role of thermal fluctuations on the stability of Au-C6DT-Au structures at the low-$v_d$ conditions.[49]

Effects of mechanical straining on electron transport through the C6DT junctions were evaluated by investigating the trace shape of plateaus at around $10^{-3}$ $G_0$ in the individual $G$–$t$ curves acquired at $v_d = 0.6$ pm/s (Figure 1c) and 6 pm/s (Figure 1d). The single-molecule conductance traces could be classified into three groups from their inclination: positive (conductance increase, CI), flat (FL), and negative (conductance decrease, CD) dependence of $G$ on $L_d$ (curves showing large fluctuations are excluded here). Our experiment thus confirms that the stretching-induced CI behavior can appear in C6DT junctions as well as BDT junctions, indicating its generality in stretched S contact-based molecular junctions.

**Correlations between strain-dependent conductance variations and contact atomic structures.** Based on the combined DFT and NEGF approach, we now establish the correlations between three distinct electromechanical characteristics of a C6DT junction and specific Au-S contact atomic structures. To model the tip of break junction electrodes with well-defined Au-S coordination numbers, we introduced one, two, and three apex Au atom(s) on top of a five-layer 4×4 Au(100) slabs (Figures 2a and 2b). We carefully checked that the pulling-induced changes in energetic, structural, and transport properties presented below are not modified by employing Au(111)-based models or a different DFT exchange-correlation functional (Supplementary Figures S1 — S4). Placing a S linker atom onto these Au apex atoms results in the well-defined S-Au CN1, CN2, and CN3, respectively. As the variations of the CN1 contact model, we also considered two monoatomic Au chain-based contact models (two-atom chain CN11 and three-atom chain CN111) and a Au atom linked to two Au apex atoms (CN21). Based on these contact motifs, we prepared various symmetric junction models such as (using the notation of CN$\alpha$-$\beta$ for the combination of CN$\alpha$ contact 1 and CN$\beta$ contact 2) CN1-1, CN2-2, CN3-3, CN21-12, and CN11-11 as well as mixed-coordination asymmetric junction models such as CN3-2, CN2-1, CN11-12, CN11-1, and CN111-1.

Next, we stretched each junction model along the surface-normal direction by retracting the fixed top and bottom Au layers and successively optimizing the junction geometry. Details of junction configurations and simulation methods are presented in Methods section and Supplementary Information. The strain-dependent conductance values in different junction models calculated with the DFT-based NEGF formalism are presented in Figure 2c. They allow us to establish the connection between three distinct stretching-induced conductance variation trends observed in the experiment, CI, FL, CD, with specific contact atomic structures: First, the CN2- or CN3-based junction models (CN2-2, CN3-3, and CN2-3) show the CI mode. Next, the CN1- or CN21-based junctions (CN1-1, CN21-12, and CN2-1) can be associated with the FL mode. Finally, with the introduction of monoatomic Au chain contacts (CN11-11, CN11-1, CN11-CN12, and CN111-1), we observe the CD mode. Note the dominance ordering of CN11/CN111 > CN1/CN21 > CN2/CN3 in mixed CN junctions: For example, the CD mode from the CN11-1 junction can be understood in terms of the presence of CN11 (in spite of the presence of CN1). For the CD mode, it should be also noted that the formation of Au chains in both contacts results in a more dramatic decrease to smaller conductance values than in the one-side Au chain case.

**The origins of different strain-dependent conductance variation types: CI, FL, and CD.** To understand the microscopic origins of sensitive conductance variations depending on contact geometries, we compared the transmission, atom projected density of states (PDOS), and local density of states (LDOS) of the symmetric CN3-3, CN1-1, and CN11-11 junctions, which are the representative cases of the CI, FL, and CD modes, respectively (Figure 3). The nature of other junction models including the above-mentioned preference ordering in mixed CN cases can be understood via their extrapolation (see Supplementary Figure S5). We observe that the energetic locations of transmission peaks (Figure 3a) coincide with those of S-induced apex Au atom(s) and C6 HOMO PDOS peaks (Figure 3b). The counterintuitive conductance increase accompanying the junction gap widening can be explained by the energetic rise of the C6DT HOMO that results from the mechanical stretching of the molecules.[25-26] Note that, although all three C6DT models exhibit stretching-induced upshifts of the HOMO levels toward $E_F$ (indicated by downward blue and red arrows), the amount of upshift in CN3-3 is notably bigger than that in CN1-1 and it is almost negligible in CN11-11. This can be understood by the fact that, as shown in Figure 3c (see also Supplementary Figure S6) the nature of the electrostatic potentials $\delta V_{es}$ associated with the charge redistributions induced by the covalent Au-S bonding,[50-52]

$$\delta\rho = \rho(\text{Au}+\text{C6DT}) - \rho(\text{Au}) - \rho(\text{C6DT}) \quad (1)$$

are very different in the three CN cases: In the unstrained starting point, $\delta V_{es}$ in CN3-3 is much flatter than that in CN1-1 and CN11-11, and their stretching-induced upshifts can be quantitatively correlated with those of the HOMO levels.

To fully understand the origins of the stretching-induced conductance variations, in addition to the classical electrostatics, we need to additionally take into account the nature of S-mediated quantum-mechanical electronic coupling between C6DT HOMO and Au electrode states in different contact atomic structures. The strength of the coupling between molecular and electrode 1/2 states (escape rate or inverse lifetime) is physically manifested by the broadening of molecular energy levels and is mathematically expressed within NEGF as the anti-Hermitian component of self-energies $\Sigma_{1/2}$ ("broadening matrix"),[53]

$$\Gamma_{1/2} = i\left(\Sigma_{1/2} - \Sigma_{1/2}^{+}\right). \quad (2)$$

To begin with, in the low strain condition ($\Delta L = 0$ Å), the energetic locations of HOMO PDOS peaks in CN3-3 (Figure 3b, upper left), CN1-1 (Figure 3b, upper center), and CN11-11 (Figure 3b, upper right) were $E_F - 1.10$ eV, $E_F - 0.59$ eV, and $E_F - 0.16$ eV, respectively, becoming higher and closer to $E_F$. We also note that their broadening becomes smaller in the order of CN3-3, CN1-1, and CN11-11. This can be again related with the above-mentioned differences in the $\delta V_{es}$ in the three CN contacts (Figure 3c).

Upon stretching the junctions by $\Delta L = 3$ Å, we find that the S-induced C6 PDOS peaks upshift to $E_F - 0.60$ eV, $E_F - 0.24$ eV, and $E_F - 0.10$ eV, respectively (Figure 3b, lower panels). Overall, the amount of pulling-induced HOMO transmission peak upshift is getting smaller and its broadening also becomes notably smaller in the order of CN3-3, CN1-1, and CN11-11. For the CN3-3 case, since the C6DT HOMO PDOS peak remains to be broad or the good molecule-electrode connectivity mediated through the CN3 contact is still maintained even after bond stretching, the shift of C6DT HOMO toward $E_F$ is significant enough to induce enhanced transmission at $E_F$ (CI mode, Figure 3a, left). For the CN2-2 case, we obtained an essentially identical behavior as the CN3-3 counterpart (Supplementary Figure S5). On the other hand, in the CN11-11 case, the amount of junction pulling-induced HOMO upshift is very small and moreover the originally narrow peak becomes even narrower, resulting in the overall decreased transmission at $E_F$ (CD mode,

Figure 3a, right). The CN1 contact is the case where the C6DT HOMO upshift resulting from junction pulling is more or less compensated by the decreasing C6DT-Au spatial connectivity or level broadening (FL mode, Figure 3a, center).

The LDOS corresponding to the C6DT HOMO peaks in the three contact models shown in Figure 3d also provide the visual explanation of the above descriptions. For the CN3-3 junction model, the HOMO states show a good connectivity through the S linker atom, three Au apex atoms, and Au electrode in the low strain regime, and this spatial delocalization is relatively well preserved even in the high strain limit (Figure 3d, left), allowing the HOMO upshift to induce CI. However, in the CN11-11 (CN1-1) junction counterpart, as shown in Figure 3D right (middle) panel, the relatively delocalized HOMO states in the low strain region become strongly (moderately) localized around the S linker atoms upon junction pulling, resulting in the disconnected Au-C6DT states and associated CD (FL) behavior.

**Experimental determination of stretching rate dependence of CI, FL, and CD types.** Examining the strain rate dependence of single molecule junction lifetime represents a powerful scheme to study the stability and breakdown mechanism of metal-molecule contacts.[35, 54-56] In Figures 4a and 4b, we present the statistics of the CI, FL, and CD data for the $v_d = 6$ pm/s and $v_d = 0.6$ pm/s cases, together with the range of their breakdown conductance values. First, regardless of the stretching speed, we found that the FL and CD modes appear much more frequently than the CI counterpart. According to the above-established connection between Au-S CNs and conduction variation types, this indicates that CN1 and CN11(1) are much more preferable than CN2 or CN3. This can be understood in view of the possible sequential conversions of CN3 or CN2 (CI) into CN1 (FL) and the eventual chain formation (CD).[18, 20, 23, 30-32, 57] Next, at the increased junction pulling speed of $v_d = 6$ pm/s, we observed that the portion of CI mode is slightly increased than in the $v_d = 0.6$ pm/s case. We have previously observed a similar trend for BDT junctions, and attributed it to the suppression of contact structure relaxation processes at high strain rate conditions.[35] Again, according to the correlations with the Au-S contact CNs established earlier, we can claim that the higher stretching rate should more likely preserve CN2 or CN3 and exhibit CI.

**Stretching-induced sequential CN conversions studied *via* ab initio MD simulations.** To rationalize the above experimental observations regarding the stretching rate dependence, we further carried out *ab initio* MD simulations using the CN3-3 model. Before discussing the MD results, we first note that the static junction pulling simulations discussed earlier (Figs 2 and 3) give the relative stability ordering of CN1 > CN3 ≈ CN2 (Figure 4c). The maximum stretching distance of the CN1-1 model $\Delta L$ = 4.2 Å was also longer than the corresponding value of $\Delta L$ = 3.0 Å in the CN3-3 and CN2-2 counterparts. Detailed examination of contact atomic structures in the CN3-3 case shows the elongation of either the longest one (Figure 4c inset) or two Au-S bonds (Supplementary Figure S2), which is in line with the protrusion of a Au surface atom after the maximal stretching of a junction model based on perfectly flat Au electrodes (without any apex atom).[39] Similarly, for CN2, one of the two Au-S bonds becomes longer than the other with the junction stretching. However, upon the continued retraction of molecular junctions, in agreement with earlier studies,[28, 32, 40, 58] the static junction pulling simulations exhibited only the breaking of the longest Au-S bond. Namely, the structural transformations in Au apex atoms and resulting CN changes could not be observed. Considering the Au(111) surface, we obtained the overall identical results as in the Au(100) case (Supplementary Figures S1 and S2). The structural transformations in the Au apex region in the CN3 and CN2 contacts are however expected in view of the well-established fact that, due to the very strong bond strength of Au in low-coordinated structures, monoatomic Au chains are easily formed.[45-47] Particularly, it is now known that Au apex atoms and chains are also easily generated upon thiol-mediated molecular pulling,[18, 20, 23, 30-32] although the extent of junction stretching and the structural details of transformation are still controversial.[18, 32, 57] The very mobile nature of these Au adatoms is also considered to be a key to understand the molecular self-assembly of thiolate-linked molecules and accompanying surface reconstructions.[22-24, 59]

To properly capture the high-CN to low-CN structural rearrangement processes in realistic environments including thermal fluctuations, we carried out DFT-based first-principles MD simulations.[23, 30-31, 59-60]. Focusing on the MD simulations involving the CN3 contact, which has more degrees of freedom in structural transformation than CN2, we first found that the currently available classical force-fields[29, 33] cannot properly reproduce the continuous CN changes in the Au tips composed of several apex atoms. Our simulation approach is also discriminated from earlier *ab initio* MD studies in several aspects (see Supplementary Information for details): In terms of the junction modeling, we explicitly adopted the CN3 contact as the initial geometry. It should be noted that the CN3 geometry is difficult to appear in unbiased junction pulling MD simulations, because they should preferentially generate the energetically more favorable CN1 and CN11 structures (Figure 4c). Next, as in earlier *ab initio* MD studies, due to the very high computational cost, the total MD time obtained in our simulations is several orders of magnitude shorter than the experimental time scale. However, we evolved the system for a much longer period of time at each gap distance (e.g. 1000 fs compared with 60 fs of Ref. 31, see Supplementary Information). Our pulling MD simulations then, as summarized in Figure 4d for the CN3-3 model, showed the structural conversions in Au apex atoms and an accompanied much longer junction elongation distance of ~ 10.8 Å that could not be observed in the static junction stretching simulations: The top and bottom CN3 are first sequentially converted into CN21, and the bottom CN21 is next converted into CN111, which eventually becomes the rupture point (see also Supplementary Movie S1 — S4). We particularly note that the full contact geometry before the junction breaking, which consists of CN11 and CN12, is in excellent agreement with the most probable contact configuration identified in a recent pioneering experiment.[21] Our MD simulations thus resolve the controversy over the nature of the observed very long junction elongation by providing atomic-scale information on the junction formation process.[18, 32, 57]

Both the energetic preference of CN1 over CN2/CN3 found in static junction pulling simulations and the possibility of CN3 (CN2) conversion into CN21 (CN11) observed in pulling MD simulations naturally explain the very small ratio of junctions that exhibited stretching-induced CI (Figure 4a). We additionally repeated the above MD simulations for the initial CN3-3 to CN21-3 conversion process by systematically increasing the pulling speed (see Supplementary Table S1). Then, at the MD evolution time of 100 fs and 120 fs, which is about an order of magnitude reduced from the original evolution time of 1000 fs, we could observe the direct junction breaking at CN3 rather than the CN3-to-CN21 conversion. This again provides a consistent atomistic explanation for the more frequent experimental observation of stretching-induced CI upon increasing the pulling speed from $v_d$ = 0.6 pm/s to $v_d$ = 6 pm/s (suppression of high-CN to low-CN structural transformations).

**Origins of multiple conductance peaks in single-molecule junctions.** We finally point out the relevance of our study for the long-standing question on the origins of ubiquitous multiple conductance peaks in S-Au contact-based single-molecule junctions. From alkanedithiolate junction experiments, two,[6-7, 15-16] three,[9-10, 14, 17] and even four [8, 12, 18-19] conductance peaks were reported, and they were usually attributed to different S-Au contact geometries except the lowest conductance values that were assigned to gauche defects [8-9, 18, 31] or thiol (rather than thiolate) molecule termination.[20, 40, 58, 61] In our experiments, we thus concentrate only on the 'high' and 'medium' conductance regimes, and do not attempt to identify the 'low' conductance peaks that appear below $10^{-4}$ $G_0$. First, by collecting all three CI, FL, and CD data sets, as typically done in previous studies, two prominent peaks emerged in the conductance histograms shown in Figure 4b (right panel). In the $v_d$ = 0.6 pm/s ($v_d$ = 6 pm/s) data set, the most pronounced first conductance peak appears at $3.5 \times 10^{-4}$ $G_0$ ($3.4 \times 10^{-4}$ $G_0$), while the second peak at $8.4 \times 10^{-4}$ $G_0$ ($7.0 \times 10^{-4}$ $G_0$). Note that the first and second peaks correspond well with the 'medium' and 'high' conductance peaks reported in the literature.[17, 21] On the other hand, combining the $v_d$ = 0.6 pm/s and $v_d$ = 6 pm/s data sets, the conductance peaks in the CD-, FL-, and CI-mode histograms appear at $3.5 \times 10^{-4}$ $G_0$, $4.7 \times 10^{-4}$ $G_0$, and $6.8 \times 10^{-4}$ $G_0$, respectively (see Supplementary Figure S7). The comparison between our new classification scheme, which provides a more detailed information of Au-S contact configurations, and the conventional one now explains the difficulty of identifying specific contact geometries based on multiple conductance peak values alone: The significant FL histogram peak (CN1/CN21 contacts) is buried by the more dominant CD peak (CN11/CN111 contacts), and cannot be identified at all within the conventional data processing method. The FL conductance data in turn slightly (strongly) contaminate the CD (CI) histograms, which can be associated with the CN11/CN111 (CN2/CN3) conformations [9, 17, 31]. It is still pleasing to note that a simple augmentation to the conventional statistical analysis scheme will be sufficient to achieve more accurate identifications of S-Au contact atomic structures.

## Summary and Conclusion

In conclusion, for the representative Au-C6DT-Au single-molecule junctions, we experimentally demonstrated the presence of stretching-induced increasing conductance (CI) mode as well as more dominant flat (FL) and decreasing (CD) counterparts, and theoretically established the correlations between the three distinct electromechanical characteristics and different Au-S contact atomic structures. Specifically, the CI, FL, and CD conductance variation types were associated with CN2/CN3, CN1/CN21, and CN11/CN111 contacts, respectively, and the microscopic origins of this classification were explained in terms of the CN-dependent upshift of the stretching-induced interface-dipole potential (in the order of CN2/CN3 > CN1/CN21 > CN11/CN111) and the decrease in molecule-electrode electronic connectivity (in the order of CN2/CN3 < CN1/CN21 < CN11/CN111). Based on experiments performed at two different pulling rates and corresponding *ab initio* MD simulations, we further clarified the critical role of continuously occurring atomic structural rearrangements at the Au electrode tip and accompanying Au-S CN conversions that include the eventual formation of monoatomic Au chains. Our findings provide a simple yet effective framework to identify contact atomic structures in single-molecule junction experiments and resolve the controversies over the origins of multiple conductance peaks[6-19] and a very long junction stretching length.[18, 32, 57] The finding of the dominance of Au monoatomic chain configurations in single-molecule junctions also indicates that caution must be exercised when one applies the knowledge accumulated for single-molecule junctions to self-assembled monolayer-based counterparts,[22-24] which are more relevant for practical device applications.

## Acknowledgements

This research was supported by the Basic Research Program (NRF-2017R1A2B3009872), the Global Frontier Program (2013M3A6B1078881), and the Nano·Material Technology Development Program (NRF-2016M3A7B4024133) of the National Research Foundation funded by the Ministry of Science, ICT, and Future Planning of Korea. Computational resources were provided by the KISTI Supercomputing Center (KSC-2016-C3-0076).

## Author Information

### Contributions
Y.-H.K. conceived the idea and oversaw the computational research. H.S.K. and J.L. performed ab initio calculations, and Y.-H.K., H.S.K., and J.L. analyzed the results. M.T. designed and conducted experiments, and M.T. and T.K. analyzed the results. Y.-H.K. and M.T. co-wrote the manuscript, with contributions from all authors.

### Competing interests
The authors declare no competing financial interests.

### Corresponding Authors
*Y.-H.K.: y.h.kim@kaist.ac.kr
*M.T.: makusu32@sanken.osaka-u.ac.jp

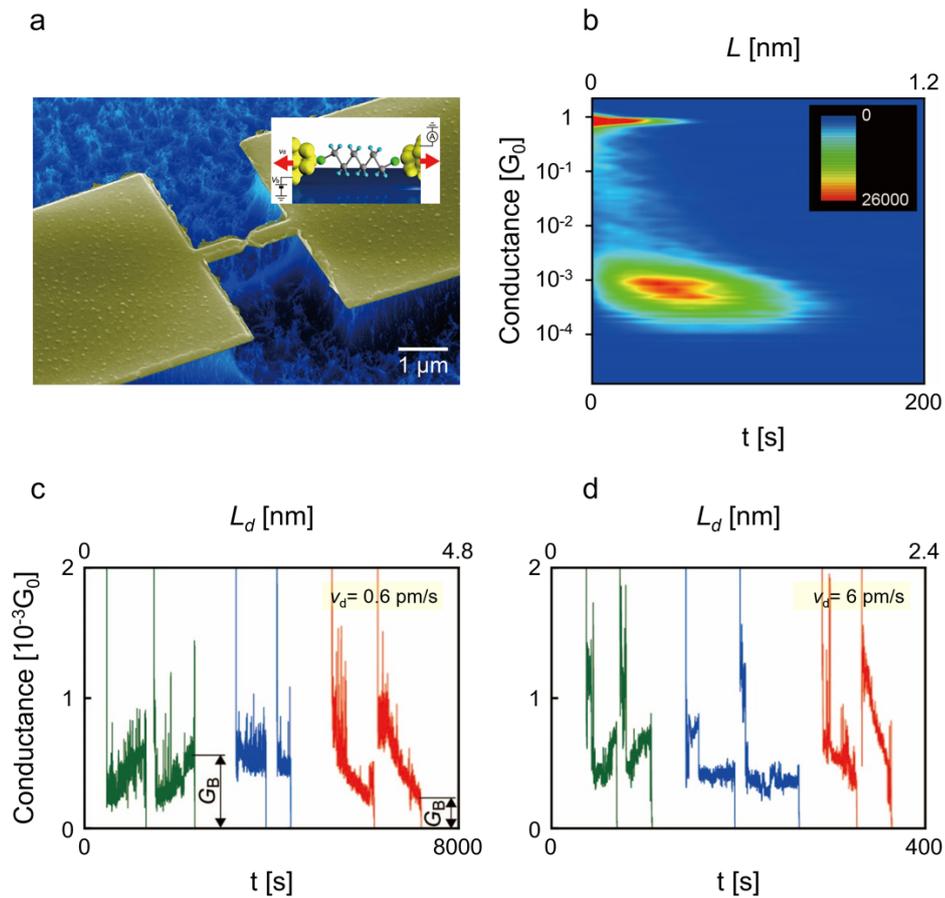

**Figure 1 | Experimental conductance measurements on mechanically-stretched Au-C6DT-Au junctions and stretching-dependent conductance of C6DT single-molecule junctions.** (**a**) A false-colored scanning electron micrograph of a micro-fabricated mechanically-controllable break junction consisting of a 2 μm-long free-standing Au wire on a polyimide-coated bending beam. A schematic model depicting single-molecule conductance measurements of a C6DT molecule connected to two Au electrodes (inset). The conductance is recorded under the applied voltage $V_b$ = 0.2 V with strain imposed to the junction through mechanical displacements of the two electrodes at a constant rate $v_d$. (**b**) A two-dimensional histogram of the C6DT junction conductance obtained under $v_d$ = 6 pm/s. Three types of single-molecule conductance versus time curves obtained under $v_d$ of 0.6 pm/s (**c**) and 6 pm/s (**d**) exhibiting positive (green), flat (blue), and negative (red) dependence on the mechanical strain imposed. $G_B$ denotes the conductance states at the point of junction breakdown.

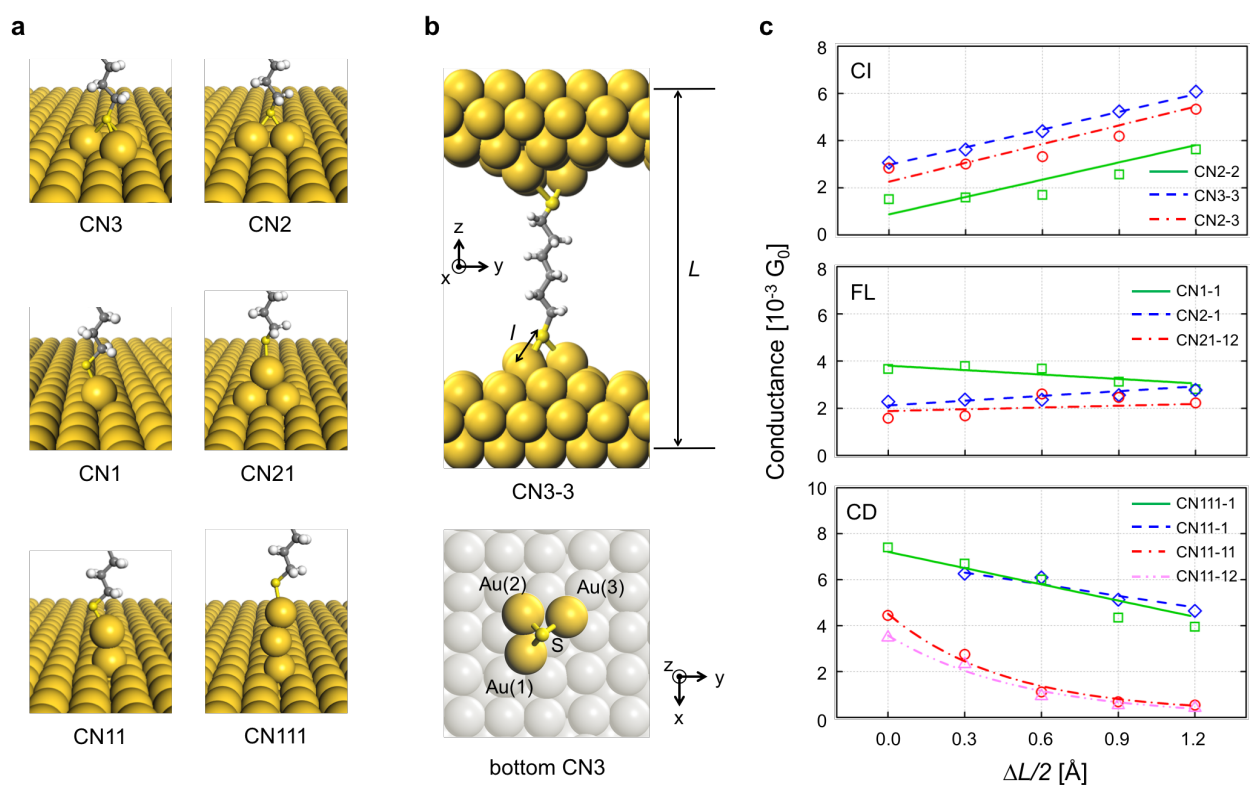

**Figure 2 | Molecule-electrode contact structural motifs that result in three different calculated pulling-induced conductance variation trends.** (a) Various contact atomic structures with different Au-S CNs can be grouped into three families according to whether they induce the CI (CN3, CN2) or FL (CN1, CN21) or CD (CN11, CN111) conductance variation behavior upon junction stretching. (b) The side (upper panel) and top views (lower panel) of a representative junction model based on the CN3 contact (CN3-3) and Au(100) electrode surface. In the top view, the atomic structure of the bottom Au-S contact is shown. Here, $L$ is defined as the distance between the top fixed layer of the bottom electrode and the bottom fixed layer of the top electrode (third layers from the electrode surface; see Supporting Information for details). The lowercase $l$ represents the Au-S bond length of three Au-S bonds. The initial distance $L_0$ ($\Delta L = 0.0$ Å) for the CN3-3 model was 23.7 Å for the CN3-3 model. (c) Calculated strain-dependent conductance variations grouped into three types: CI (top panel), FL (middle panel), and CD (bottom panel). Lines are added to guide the eye.

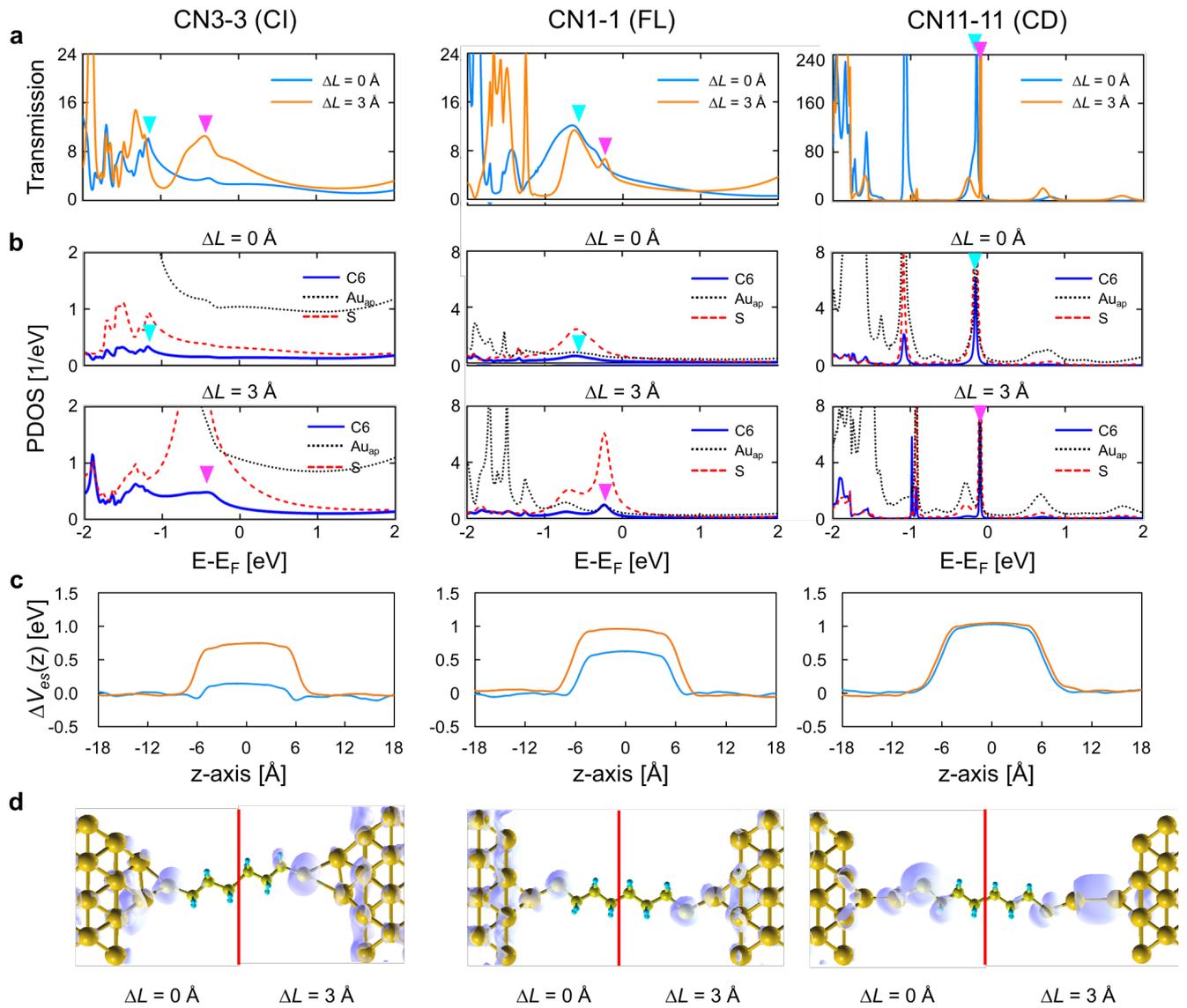

**Figure 3 | Microscopic origins of three different conductance variation types.** (**a**) Transmissions of the CN3-3, CN1-1, and CN11-11 models at the low (cyan lines) and high (orange lines) strain conditions, and (**b**) the corresponding PDOS plots at the low (upper panel) and high (lower panel) strain conditions. In (b), blue solid, black dotted, and red dashed lines represent the hexane core, Au apex atoms, and S linker PDOS, respectively. Downward triangles in (a) and (b) indicate the S-originated HOMO-induced transmission peak positions. The corresponding (**c**) electrostatic potentials induced by metal-molecule charge transfers at the low (cyan lines) and high (orange lines) strain conditions, and (**d**) the LDOS around the HOMO-induced transmission peak positions with the energy window of [-0.05, +0.05] eV at the low (left panel) and high (right panel) strain conditions

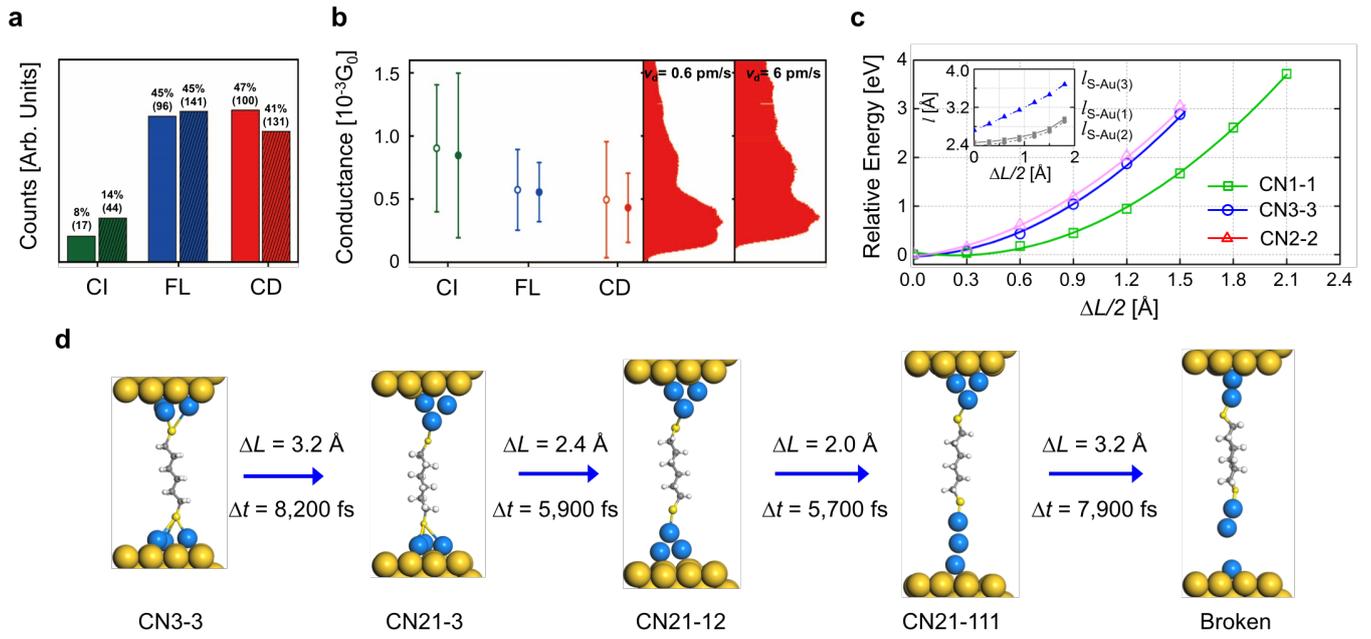

**Figure 4 | Changes in the measured statistics depending on the pulling speed and the computed energetics and dynamics of Au-S contacts.** (**a**) Statistics of the CI, FL, and CD groups from break junction experiments. Filled and shaded bars represent the number of events detected under $v_d$ = 0.6 pm/s and 6 pm/s, respectively. (**b**) Breakdown conductance $G_B$ for the three types of events under $v_d$ = 0.6 pm/s (open circles) and 6 pm/s (filled circles). Error bars are the standard deviation. Shown on the right are the conductance histograms obtained by compiling the CI, FL, and CD datasets. (**c**) Stretching-induced relative energy variations of the CN1-1, CN2-2, and CN3-3 models as the function of displacement ($\Delta L$) within the static junction elongation simulations. Inset shows the corresponding variations of the three Au-S bond lengths for the bottom contact of the CN3-3 model (For the atom labels, refer to Figure 2b lower panel). (**d**) Representative ab initio MD snapshots for the high- to low-CN changes (CN3 → CN21 in both contacts) that include the eventual Au monoatomic chain formation (CN21 → CN111 in the top contact) and junction rupture (see also Supplementary Movie S1 — S4).